\documentstyle[twoside]{article}
\input{epsf}
\catcode`\@=11
\long\def\@makefntext#1{
\protect\noindent \hbox to 3.2pt {\hskip-.9pt  
$^{{\eightrm\@thefnmark}}$\hfil}#1\hfill}		

\def\@makefnmark{\hbox to 0pt{$^{\@thefnmark}$\hss}}	
	
\def\ps@myheadings{\let\@mkboth\@gobbletwo
\def\@oddhead{\hbox{}
\rightmark\hfil\eightrm\thepage}   
\def\@oddfoot{}\def\@evenhead{\eightrm\thepage\hfil
\leftmark\hbox{}}\def\@evenfoot{}
\def\sectionmark##1{}\def\subsectionmark##1{}}

\oddsidemargin=\evensidemargin
\newcounter{sectionc}\newcounter{subsectionc}\newcounter{subsubsectionc}
\renewcommand{\section}[1] {\vspace{12pt}\addtocounter{sectionc}{1} 
\setcounter{subsectionc}{0}\setcounter{subsubsectionc}{0}\noindent 
	{\tenbf\thesectionc. #1}\par\vspace{5pt}}
\renewcommand{\subsection}[1] {\vspace{12pt}\addtocounter{subsectionc}{1} 
	\setcounter{subsubsectionc}{0}\noindent 
	{\bf\thesectionc.\thesubsectionc. {\kern1pt \bfit #1}}\par\vspace{5pt}}
\renewcommand{\subsubsection}[1] {\vspace{12pt}\addtocounter{subsubsectionc}{1}
	\noindent{\tenrm\thesectionc.\thesubsectionc.\thesubsubsectionc.
	{\kern1pt \tenit #1}}\par\vspace{5pt}}
\newcommand{\nonumsection}[1] {\vspace{12pt}\noindent{\tenbf #1}
	\par\vspace{5pt}}

\newcounter{appendixc}
\newcounter{subappendixc}[appendixc]
\newcounter{subsubappendixc}[subappendixc]
\renewcommand{\thesubappendixc}{\Alph{appendixc}.\arabic{subappendixc}}
\renewcommand{\thesubsubappendixc}
	{\Alph{appendixc}.\arabic{subappendixc}.\arabic{subsubappendixc}}

\renewcommand{\appendix}[1] {\vspace{12pt}
        \refstepcounter{appendixc}
        \setcounter{figure}{0}
        \setcounter{table}{0}
        \setcounter{lemma}{0}
        \setcounter{theorem}{0}
        \setcounter{corollary}{0}
        \setcounter{definition}{0}
        \setcounter{equation}{0}
        \renewcommand{\thefigure}{\Alph{appendixc}.\arabic{figure}}
        \renewcommand{\thetable}{\Alph{appendixc}.\arabic{table}}
        \renewcommand{\theappendixc}{\Alph{appendixc}}
        \renewcommand{\thelemma}{\Alph{appendixc}.\arabic{lemma}}
        \renewcommand{\thetheorem}{\Alph{appendixc}.\arabic{theorem}}
        \renewcommand{\thedefinition}{\Alph{appendixc}.\arabic{definition}}
        \renewcommand{\thecorollary}{\Alph{appendixc}.\arabic{corollary}}
        \renewcommand{\theequation}{\Alph{appendixc}.\arabic{equation}}
        \noindent{\tenbf Appendix \theappendixc #1}\par\vspace{5pt}}
\newcommand{\subappendix}[1] {\vspace{12pt}
        \refstepcounter{subappendixc}
        \noindent{\bf Appendix \thesubappendixc. {\kern1pt \bfit #1}}
	\par\vspace{5pt}}
\newcommand{\subsubappendix}[1] {\vspace{12pt}
        \refstepcounter{subsubappendixc}
        \noindent{\rm Appendix \thesubsubappendixc. {\kern1pt \tenit #1}}
	\par\vspace{5pt}}

\newcommand{\textlineskip}{\baselineskip=13pt}
\newcommand{\smalllineskip}{\baselineskip=10pt}

\def\eightcirc{
\begin{picture}(0,0)
\put(4.4,1.8){\circle{6.5}}
\end{picture}}
\def\eightcopyright{\eightcirc\kern2.7pt\hbox{\eightrm c}}

\def\abstracts#1#2#3{{
	\centering{\begin{minipage}{4.5in}\baselineskip=10pt\footnotesize
	\parindent=0pt #1\par 
	\parindent=15pt #2\par
	\parindent=15pt #3
	\end{minipage}}\par}} 

\renewenvironment{thebibliography}[1]
	{\frenchspacing
	 \ninerm\baselineskip=11pt
	 \begin{list}{\arabic{enumi}.}
	{\usecounter{enumi}\setlength{\parsep}{0pt}
	 \setlength{\leftmargin 12.7pt}{\rightmargin 0pt} 
	 \setlength{\itemsep}{0pt} \settowidth
	{\labelwidth}{#1.}\sloppy}}{\end{list}}

\newcounter{itemlistc}
\newcounter{romanlistc}
\newcounter{alphlistc}
\newcounter{arabiclistc}

\newcommand{\fcaption}[1]{
        \refstepcounter{figure}
        \setbox\@tempboxa = \hbox{\footnotesize Fig.~\thefigure. #1}
        \ifdim \wd\@tempboxa > 5in
           {\begin{center}
        \parbox{5in}{\footnotesize\smalllineskip Fig.~\thefigure. #1}
            \end{center}}
        \else
             {\begin{center}
             {\footnotesize Fig.~\thefigure. #1}
              \end{center}}
        \fi}

\newcommand{\tcaption}[1]{
        \refstepcounter{table}
        \setbox\@tempboxa = \hbox{\footnotesize Table~\thetable. #1}
        \ifdim \wd\@tempboxa > 5in
           {\begin{center}
        \parbox{5in}{\footnotesize\smalllineskip Table~\thetable. #1}
            \end{center}}
        \else
             {\begin{center}
             {\footnotesize Table~\thetable. #1}
              \end{center}}
        \fi}

\def\@citex[#1]#2{\if@filesw\immediate\write\@auxout
	{\string\citation{#2}}\fi
\def\@citea{}\@cite{\@for\@citeb:=#2\do
	{\@citea\def\@citea{,}\@ifundefined
	{b@\@citeb}{{\bf ?}\@warning
	{Citation `\@citeb' on page \thepage \space undefined}}
	{\csname b@\@citeb\endcsname}}}{#1}}

\newif\if@cghi
\def\cite{\@cghitrue\@ifnextchar [{\@tempswatrue
	\@citex}{\@tempswafalse\@citex[]}}
\def\citelow{\@cghifalse\@ifnextchar [{\@tempswatrue
	\@citex}{\@tempswafalse\@citex[]}}
\def\@cite#1#2{{$\null^{#1}$\if@tempswa\typeout
	{IJCGA warning: optional citation argument 
	ignored: `#2'} \fi}}

\def\pmb#1{\setbox0=\hbox{#1}
	\kern-.025em\copy0\kern-\wd0
	\kern.05em\copy0\kern-\wd0
	\kern-.025em\raise.0433em\box0}

\def\fnt#1#2{\footnotetext{\kern-.3em
	{$^{\mbox{\scriptsize #1}}$}{#2}}}

\def\fpage#1{\begingroup
\voffset=.3in
\thispagestyle{empty}\begin{table}[b]\centerline{\footnotesize #1}
	\end{table}\endgroup}

\def\runninghead#1#2{\pagestyle{myheadings}
\markboth{{\protect\footnotesize\it{\quad #1}}\hfill}
{\hfill{\protect\footnotesize\it{#2\quad}}}}
\font\tenrm=cmr10
\font\tenit=cmti10 
\font\tenbf=cmbx10
\font\bfit=cmbxti10 at 10pt
\font\ninerm=cmr9

\font\eightrm=cmr8

\def\qed{\hbox{${\vcenter{\vbox{			
   \hrule height 0.4pt\hbox{\vrule width 0.4pt height 6pt
   \kern5pt\vrule width 0.4pt}\hrule height 0.4pt}}}$}}

\begin{document}
\catcode`@=11
\def\@cite#1{${}^{\mbox{\small #1}}$}
\catcode`@=11

\runninghead{ Production of the $B_c$ Meson } 
            { Production of the $B_c$ Meson }

\normalsize\textlineskip
\thispagestyle{empty}
\setcounter{page}{1}
\hfill \vbox{\halign{&#\hfil\cr
		& OHSTPY-HEP-T-96-029	\cr
		& September 1996	\cr}}
\vspace*{0.44truein}

\fpage{1}
\centerline{\bf PRODUCTION OF THE $B_c$ MESON\footnote{
 Invited talk presented at the Quarkonium Physics Workshop, 
 University of Illinois at Chicago, June 1996.
}}
\vspace*{0.37truein}
\centerline{\footnotesize YU-QI CHEN }
\vspace*{0.015truein}
\centerline{\footnotesize\it  Department of Physics, Ohio State University }
\baselineskip=10pt
\centerline{\footnotesize\it Columbus, OH 43210, USA }

\vspace*{0.21truein}
\abstracts{The production of the $B_c$ and $B_c^*$ mesons was
studied in the framework of the factorization formalism and 
 perturbative QCD.  Predicted results implied that an observable 
number of $B_c$ and $B_c^*$ events can be produced at LEP I and at Tevatron.  
The fragmentation  approximation describes the production of the  $B_c$ 
and $B_c^*$ mesons very well  in high energy $e^+e^-$ collisions. 
In hadronic collisions, it is valid when and only when the transverse 
momentum $P_T$ of the produced $B_c$ and $B_c^*$ is much larger than the 
mass of the $B_c$ meson.  
}{}{}


\vspace*{1pt}\textlineskip	
\section{Introduction }	        
\vspace*{-0.5pt}
\noindent
The $B_c$ meson is  very interesting because it is the only flavored meson
containing a heavy quark and a heavy antiquark. 
 However, the study of the $B_c$ meson
suffers from the difficulty of producing it in experiments. The conventional
production mechanisms for the $B_u$ and $B_d$ mesons and for heavy quarkonia
$J/\psi$ and $\Upsilon$ are not available for  the
$B_c$ and $B_c^*$ mesons. A question is  whether a sufficiently large number of $B_c$
events can  be produced in experiments. If not, the study of the $B_c$ 
meson would be only of theoretical interest. 
  
 Theoretical studies of   the production for the $B_c$ meson over 
the past few years give a positive answer to this question. 
The dominant mechanism 
for the production of the $B_c$ meson in high energy collisions has been
found to be the following. First 
two heavy quark pairs $c\bar{c} $ and $ b \bar{b} $ are produced
in a high energy process, and then the  
$c$ quark from  the $ c\bar{c} $ pair
and the $\bar{b}$ quark from the $b\bar{b}$ pair bind to form a $B_c$ meson. 
At the $e^+e^-$ collider LEP I,
the $B_c $ meson  can be produced via the $Z^0$ decay\cite{9,60}
$
 e^+e^-\to Z^0\rightarrow B_c +b+\bar{c}.
$
At the hadron-hadron colliders Tevatron and LHC, it can be produced by  
gluon-gluon fusion and quark-antiquark annihilation subprocesses\cite{0} 
such as
$
g+g\rightarrow B_c +b+\bar{c},
$
$
q+\bar{q}\rightarrow B_c +b+\bar{c} .
$
Calculations  show that  quite a large number of the $B_c$ 
events can be produced both at LEP I and 
at the Tevatron and the LHC. The experimental search for the $B_c$
meson is now under way at LEP and the Tevatron. Some preliminary results from
LEP I experiments were  
reported in this workshop\cite{Bc-LEP}.  Results from the Tevatron are expected 
in the near future.

Like the $c\bar{c}$ and $b\bar{b}$ systems, the $\bar{b}c$  is a
nonrelativistic bound state system of a heavy quark and a heavy antiquark.
According to the nonrelativistic QCD (NRQCD) factorization 
formalism\cite{B-B-L}, the 
production cross  section of the $B_c$ and the $B_c^*$ mesons
can be factorized as the products of 
short distance coefficients and long distance matrix elements.
The production of the $c\bar{c}$ and the $b\bar{b}$ pairs is a short-distance
process that occurs  at a distance scale of order  $1/m_Q$ or smaller, 
and it can be 
accurately calculated using perturbative QCD to any order in $\alpha_s(m_Q)$. 
The formation of the 
$B_c$ meson is a long-distance process which occurs at a scale of order 
of $1/(m_Q v )$, and it can be 
described by  NRQCD matrix elements
which scale in a definite way with $v$, the typical relative 
velocity of the bound states.  In the 
production of the $B_c$ or $B_c^*$ mesons,  the contribution from
 color-octet  processes is smaller than that from color-singlet process
simply because the short-distance coefficients are of the same order 
in $\alpha_s$,
but the color-octet matrix elements are   higher order of $v$. Thus
the  production is 
dominated by  color-singlet processes.   The relevant  color-singlet matrix 
elements can be estimated well by  heavy-quark potential models or by lattice
calculations. Therefore, the production of the $B_c$ is  predictable
without any free parameters.

If  the $B_c$ is produced with sufficiently large transverse
momentum, then the fragmentation approximation is expected 
to be available. In this approximation,
the cross section factors into a cross section for producting a $b$ 
quark and a fragmentation function that gives the probability for the $b$ quark 
to hadronize into a $B_c$ meson.
The fragmentation functions 
are universal, so they  can be used 
to calculate the production of the $B_c$ both in $e^+e^-$ process
and in hadron-hadron collisions. 
An interesting question is how well the fragmentation 
approximation to the full  calculation works for the 
production of the $B_c$ and $B_c^*$ mesons.
In the following sections, this question will be addressed for the 
production of $B_c$ and $B_c^*$ at LEP and at hadronic colliders.

\vspace*{1pt}\textlineskip	
\section{ Production of the $B_c$ Meson in $e^+e^-$ Collision } 
\vspace*{-0.5pt}
\noindent
At LEP I, millions of  $Z^0$ events have been accumulated. Such a
large number of  events provides us with an opportunity to search for the 
$B_c$ meson through the process
$
Z^0 \to B_c (B_c^*) + b + \bar{c} .
$
Calculations of the  production rate\cite{8,9,60} show that 
the branching ratio for $Z^0 \to B_c + X $ is  
around $10^{-4}$.

The process can be calculated either using the covariant projection 
method\cite{K-K-S} or using the recently developed threshold expansion 
method\cite{Braaten-Chen}. At the lowest order in $\alpha_s$, there 
are four Feynman 
diagrams responsible for the process $Z^0 \to B_c +b + \bar{c} $. 
Let $k$, $P$, $q_1$, and $q_2$ be the momenta of the $Z^0$ boson, $B_c$ meson, 
$\bar{c}$ quark and $b$ quark, respectively. We can define  three Lorentz
invariant variables
\begin{equation}
x \equiv {2 k \cdot q_1 \over M_Z^2 } ,  ~~~~~
y \equiv {2 k \cdot q_2 \over M_Z^2 } ,  ~~~~~
z \equiv {2 k \cdot P \over M_Z^2 } , 
\end{equation}
which are  twice the energy fractions of the $B_c$, 
$\bar{c}$ quark  and $b$ quark
in the $Z^0$ rest frame, respectively.  They are constrained by 
$ x + y + z = 2 $ and $ 0 < x,y,z <  1 $.

The full calculation for
 $Z^0 \to B_c + X $  is straightforward. The decay width and various 
distributions in $x$, $y$, and $z$  can be expressed as products of 
short-distance coefficients and  long-distance color-singlet 
matrix elements
$\langle 0 | O_1^{B_c }  (^1S_0) | 0 \rangle $ and
$\langle 0 | O_1^{B_c^*} (^3S_1) | 0 \rangle $. 
Using the approximate heavy-quark spin symmetry of NRQCD, 
these two matrix elements are equal to $f_{B_c}^2 m_c$ and 
$3 f_{B_c}^2  m_c$, respectively,  up to corrections of
order $v^2$.

The fragmentation functions for the production of the $B_c$ can be
extracted by taking the limit $ M_Z/ m_{B_c} \to \infty $. In this limit,
the square of the matrix element is singular when the produced 
$B_c$ is  collinear
either with the $b$ quark or with the $\bar{c}$ quark. In this region, 
either $x \to 1 $ or $ y \to 1 $ but $z$ varies from  0 to 1.
In the limit, the $z$ distribution comes from 
those most singular terms and can be factorized as 
\begin{equation}
{d \Gamma_{Z^0 \to B_c^{(*)} } \over d z } = \Gamma (z \to b \bar{b} ) \cdot
                     D_{\bar{b} \to B_c^{(*)} } (z) +
\Gamma (z \to c \bar{c} ) \cdot D_{c \to B_c^{(*)} } (z) ,
\label{Z-Bc}
\end{equation}
where $D_{\bar{b} \to B_c^{(*)} } (z)$ and $ D_{c \to B_c^{(*)} } (z) $ 
are the fragmentation functions for $ \bar{b} $ or $ c $  into 
a $ B_c^{(*)} $ meson.  The $\bar{b}$ quark fragmentation functions 
 are  
\begin{eqnarray}
D_{{\bar b} \rightarrow B_c}(z)
&=& {4 \alpha_s(2 m_c)^2 \langle 0 | O_1^{B_c }  (^1S_0) | 0 \rangle 
  \over 243 m_c^3}
\, {r z (1-z)^2 \over (1 - (1-r)z)^6}
\Bigg(6 - 18(1-2r)z + (21
\nonumber \\
& & \hspace{-0.9in}
-74r+68r^2)z^2 - 2(1-r)(6-19r+18r^2)z^3 + 3(1-r)^2(1-2r+2r^2)z^4 \Bigg) \,,
\label{DB} \end{eqnarray}
for the $B_c$ meson and 
\begin{eqnarray}
D_{{\bar b} \rightarrow B_c^*}(z)
&=& {4 \alpha_s(2 m_c)^2 \langle 0 | O_1^{B_c^*} (^3S_1) | 0 \rangle  
\over 243 m_c^3} \,
{r z (1-z)^2 \over (1 - (1-r) z)^6}
\Bigg(2 - 2(3-2r)z + 3(3
\nonumber \\
& & \hspace{-0.9in}
-2r +4r^2)z^2 - 2(1-r)(4-r+2r^2)z^3
	+ (1-r)^2(3-2r+2r^2)z^4 \Bigg) \, ,
\label{DBstar} 
\end{eqnarray}
for the $B_c^*$ meson, where $r = m_c/(m_b+m_c)$. 
The $c$ quark fragmentation functions can be obtained by exchanging 
$m_b$ and $m_c$. 
These fragmentation functions describe the 
probability for a jet initiated by a high energy $\bar{b}$ quark to
include the hadron $B_c$ carrying a fraction $z$ of the jet momentum.
They were first calculated by Chang and Chen\cite{9} and Braaten, Cheung, and 
Yuan\cite{60}. 
These are  the lowest order fragmentation functions
at the energy scale around $m_{B_c}$. 
All leading logarithmic terms  
coming from the  collinear emission of  gluons 
 can be summed up by solving the 
 Altarelli-Parisi evolution 
equations  from the energy scale at order  $M_Z$ 
to that at order  $m_{B_c}$\cite{60}
\begin{equation} 
{
\mu {\partial \ \over \partial \mu} D_{i \rightarrow B_c}(z,\mu)
\;=\; \sum_j \int_z^1 {dy \over y} \; P_{i\rightarrow j}(z/y,\mu)
	\; D_{j \rightarrow B_c}(y,\mu) \;,
} 
\label{evol} 
\end{equation}
where $P_{i\rightarrow j}(x,\mu)$ is the Altarelli-Parisi
function for the splitting of the parton of type $i$ into a parton of
type $j$ with longitudinal momentum fraction $x$.

Explicit calculations show that
the result predicted by the fragmentation approximation 
Eq.~(\ref{Z-Bc}) is very close to that from 
the full calculation\cite{9,60}.

Integrating over the variable $z$, in eqs.~(\ref{DB}),~(\ref{DBstar}),
the probabilities for a high energy $\bar{b}$ antiquark or $c$ quark
fragmentating into a $B_c$ or $B_c^*$ meson can be obtained.
Taking the parameters adopted  by Braaten, Cheung, and Yuan\cite{60},
 the numerical values for the fragmentation probabilities
are $2.2 \times 10^{-4}$ for $\bar b \rightarrow B_c$ and
$3.1 \times 10^{-4}$ for $\bar b \rightarrow B_c^*$.
The fragmentation probabilities for $c$
are smaller than those for $\bar b$ by almost
two orders of magnitude, because it is much harder to creat
an extra $b\bar{b}$ pair by emitting a virtual gluon
 than to creat an extra $c\bar{c}$ pair.

\vspace*{1pt}\textlineskip	
\section{ Production of the $B_c$ Meson in Hadron-Hadron  Collision } 
\vspace*{-0.5pt}
\noindent
Hadronic production, as first pointed out by Chang and Chen\cite{0}, is 
dominated at high energies by the subprocess 
$g + g\to B_c(B_c^*) + b + \bar{c}$. It can be calculated
fully to order $\alpha_s^4$ in PQCD. Since  a difficult numerical 
calculation is involved, confirmation from some other independent
calculations are necessary. Since Chang and Chen\cite{0} first presented 
the numerical results for the hadronic production, calculations have 
also been done by several other authors\cite{1,2,3,6,5,50,7}, not all 
of which are in agreement. Slabospitsky\cite{1} claimed an order of 
magnitude larger result than Chang and Chen\cite{0}. Berezhnoy
{\it   et al.}\cite{2} 
obtained a result larger than Chang and Chen\cite{0} and smaller than 
Slabospitsky\cite{1}. Masetti and Sartogo\cite{3} found a result similar 
to Berezhnoy {\it   et al.}\cite{2}. More recently, 
Berezhnoy {\it et al.}\cite{5} found that a color factor 
$\displaystyle{1\over 3}$ was overlooked 
in their previous work\cite{2}. After including this factor, their revised 
result is in agreement with Chang and Chen\cite{0}. Baranov\cite{50} 
independently also obtained results similar to Chang and Chen\cite{0}. 
Ko{\l}odziej {\it  et al.}\cite{7} presented results using different 
parton distribution functions and energy scale from that used by Chang 
and Chen\cite{0} so it difficult to directly compare their final results. 
However, their results for the cross sections for the parton 
subprocess are similar 
to others\cite{6,5}. Ko{\l}odziej {\it  et al.}\cite{7} pointed out that the 
cross section for the subprocess  calculated by Berezhnoy {\it et al.}\cite{5} 
for $\sqrt{\hat{s}} = 1 $ TeV  was incorrect. 
 The error probably came from the numerical calculation of the subprocess
 cross section since the 
 the square of the matrix element is highly singular at such a high energy.
It  would not affect  the cross section from  hadron collision 
at Tevatron or LHC energies  because 
the contributions from gluons with 
such a high energy center-of-mass are negligible. 
Therefore, we are confident that the results of the 
original  order-$\alpha_s^4$ PQCD calculation by Chang and Chen\cite{0}
and the more recent ones\cite{6,5,50,7} are now
in agreement.

An alternative way to calculate the hadronic production is to use the 
fragmentation approximation. From  general factorization theorems
it is clear that, for sufficiently large transverse momentum $P_T$ of 
the $B_c$ or $B_c^*$, 
hadronic production is dominated by fragmentation. The calculation 
can then be considerably simplified using this approximation, as was 
first done by Cheung\cite{10}. Subsequently, the comparison between 
the full $\alpha_s^4$ calculation and the fragmentation approximation
has been discussed by several authors\cite{6,5,7}. However, very different 
conclusions have been drawn, although their numerical results are 
similar. Chang {\it et al.}\cite{6} found that when $\sqrt{\hat{s}}$ and 
$p_T$ are small the difference between the fragmentation
approximation and  the $\alpha_s^4$ calculation is large. 
Berezhony {\it  et al.}\cite{5} claimed that the fragmentation approximation 
breaks down even for very large $P_T$ by examining the ratio of $B_c^*$ 
to $B_c$ production. Ko{\l}odziej {\it  et al.}\cite{7} claimed that 
the fragmentation approximation works well if $P_T$ exceeds about $5-10$ GeV, 
which is comparable to the $B_c$ mass, by investigating the $P_T$ distribution
of the $B_c$ meson.  

It is nontrivial to clarify these points since the 
full calculation is quite complicated, with 
the dominant subprocess involving 36
Feynman diagrams, but the importance of investigating this issue is twofold.
From the theoretical perspective, it provides an ideal example 
to quantitatively examine how well the fragmentation approximation 
works for calculating the hadronic production of heavy flavored 
mesons. In this process both the full $\alpha_s^4$ contributions and the 
fragmentation approximation can each be calculated reliably. 
Experimentally, it is important to have a better understanding of the 
production of the $B_c$ mesons in the small $P_T$ region where the $B_c$ 
production cross section is the largest. The $P_T$ distribution 
decreases very rapidly as $P_T$ increases.

Recently, Chang, Chen and Oakes\cite{C-C-O} carried out a detailed 
comparative study of the fragmentation 
approximation and the full $\alpha_s^4$ QCD perturbative calculation. 
We first studied the subprocesses carefully. By analyzing the 
singularities appearing in the amplitudes and in the $P_T$ distributions 
of the subprocesses, we gained some important insight into the processes. 
We then calculated the entire hadronic production cross section for 
$p\bar{p}$ (or $pp$)  collisions at the Tevatron energy. To facilitate 
a quantitative comparison we found it very instructive to examine the 
distribution in variable $z$, defined to be 
twice the energy fraction carried by the $B_c$
meson in the center of mass of the subprocess, which is in principle
observable. Investigating the $z$ distribution we found that the 
fragmentation contribution dominates when and only when $P_T\gg M_{B_c}$. 
It is insufficient to evaluate the validity of the 
fragmentation approximation by only examining the $P_T$ distribution. 

In order to obtain some insight into the process, we first focus the 
discussion on the subprocess. At the lowest order, $\alpha_s^4$,
there are 36 Feynman diagrams responsible for the dominant gluon fusion
subprocess $g+g\to B_c +b +\bar{c}$.
When the energy in the center of mass system, $\sqrt{\hat{s}}$, is much 
larger than the heavy quark mass, the main contributions to the cross
section come from the kinematic region where certain of the amplitudes
in the matrix element are nearly singular; i.e., some of the quark lines 
or gluon lines are nearly on-shell.
 In the large $P_T$ region, 
the only singularity is the collinear one for which the $B_c$ is collinear 
 with fragmenting $\bar{b}$ antiquark or collinear with the fragmenting
 $c$ quark. Therefore, at sufficiently large $P_T$,  the subprocess is 
dominated by  fragmentation. {\em However, when the  $P_T$ of 
the $B_c$ meson is small the produced $B_c$, as well as the $b$ and the 
$\bar{c}$ quarks, can be soft or collinear with the beam.} In this region 
the square of the matrix element is highly singular because  {\em two or more} 
of the internal quarks or gluons in certain Feynman diagrams can 
{\em simultaneously} be nearly on-mass-shell. Although this region is a 
smaller part of the phase space, these nearly singular Feynman diagrams, in 
fact, make a large contribution to the cross section and dominate the small 
$P_T$ region.  
The contributions to the cross sections  can therefore be 
 decomposed into two components: 
the fragmentation contribution, which dominates in the 
large $P_T$ region and the non-fragmentation component, which  dominates  
at smaller $P_T$. The contributions of these two components are 
quite clearly distinguishable in the $P_T$ distribution of the subprocess, 
particularly at large $\sqrt{\hat s}$.  In Fig. 1 we show the $P_T$ 
distribution of the subprocess when $\sqrt{\hat{s}}=200$ GeV for both 
the full $\alpha_s^4$ calculation and the fragmentation approximation
with $\alpha_s=0.2$, $m_b=4.9$ GeV, $m_c=1.5$ GeV, and $f_{B_c}=480$ MeV.
 It is easy to see in Fig. 1 that 
when $P_T$ is larger than about 30 GeV for the $B_c$ and about 40 GeV for 
the $B_c^*$, the fragmentation approximation is close to the full 
$\alpha_s^4$ calculation. 
\begin{figure}[htbp]
\vspace*{13pt}
\begin{center}
\leavevmode
\epsfysize=2.5 truein
\epsfbox{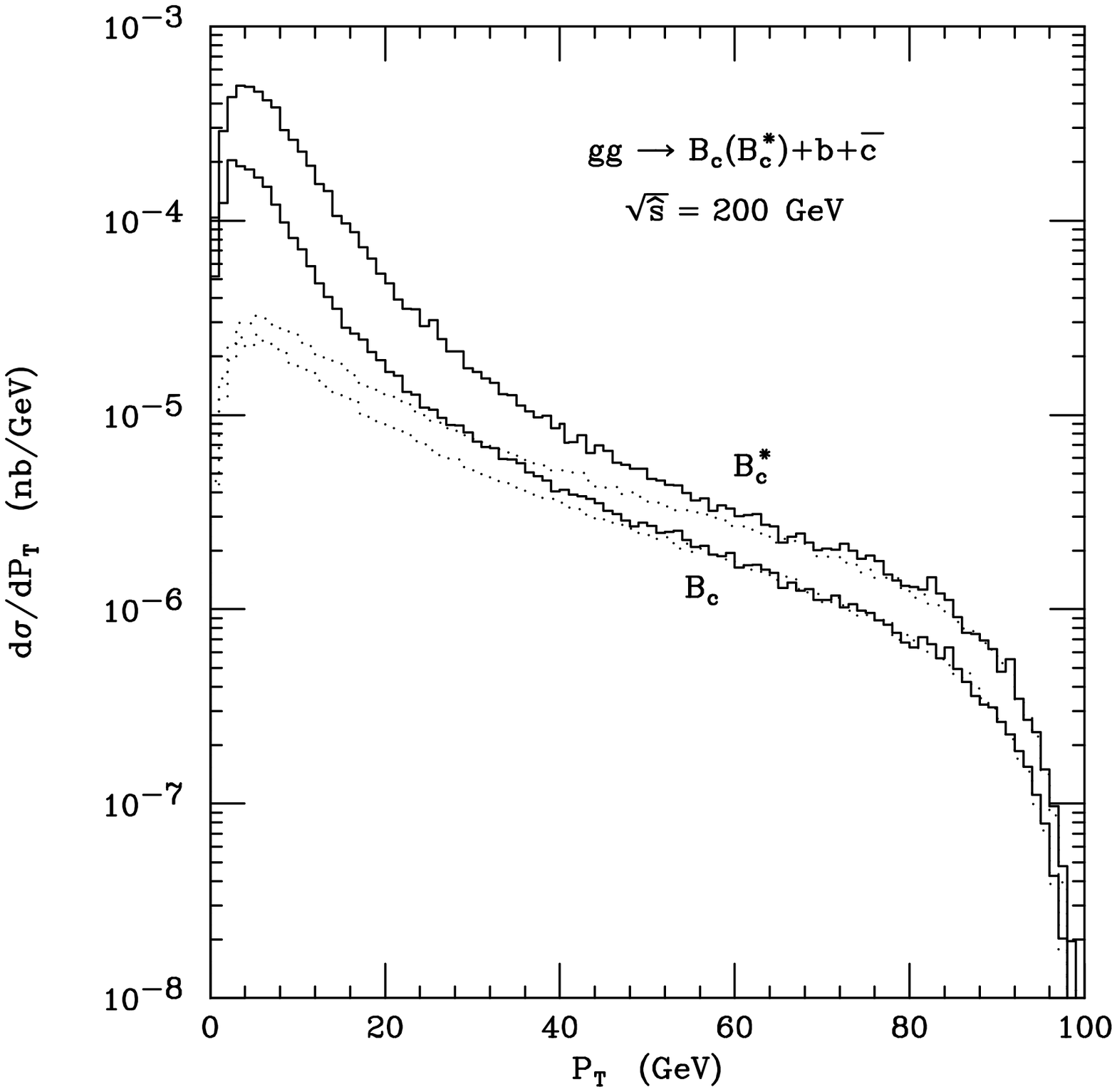}
\end{center}
\vspace*{13pt}
\fcaption{The $P_T$ distributions of the $B_c$ and $B_c^*$ meson for 
the subprocess with $\sqrt{\hat{s}}=$ 200 GeV. The solid and the doted 
lines correspond to the full $\alpha_s^4$ calculation and the fragmentation 
approximation, respectively.}
\end{figure}
However, when  the $P_T$ is smaller 
than these values,  the 
deviation between the fragmentation approximation and the full calculation 
becomes large and the non-fragmentation component clearly dominates the 
production. This critical value of $P_T$ is certainly much larger than the 
heavy quark masses, or the $B_c$ meson mass;  it slowly 
increases with increasing $\hat{s}$, which may indicate that there is an 
additional enhancement due to logarithmic terms such as $\ln{\hat{s}/m^2}$ 
in the non-fragmentation component compared to the fragmentation component. 
When ${\sqrt{\hat s}}$ is not very large this two component decomposition is 
less distinct, since the higher twist terms can not be ignored\cite{6,5,7}. 
In this case, the fragmentation approximation is not a very good 
approximation\cite{6,5,7}. 

Let us now  turn to the calculation of $B_c$ and $B_c^*$
production at $p\bar{p}$ colliders, particularly the Fermilab Tevatron. 
In Fig. 2, the $P_T$ distributions of the $B_c$ 
and the $B_c^*$ mesons coming from the full $\alpha_s^4$ calculation is 
compared with
the fragmentation approximation. In these calculations we used the CTEQ3M 
parton distribution functions\cite{16}.  
For 
simplicity, we  use a uniform choice of energy scales,  
choosing the same scale as set in  the $\bar{b}$ fragmentation
component; i.e. $\alpha_s^2(P_{T })\times \alpha_s^2(2 m_c ) $.
\begin{figure}[htbp]
\vspace*{13pt}
\begin{center}
\leavevmode
\epsfysize=2.5 truein
\epsfbox{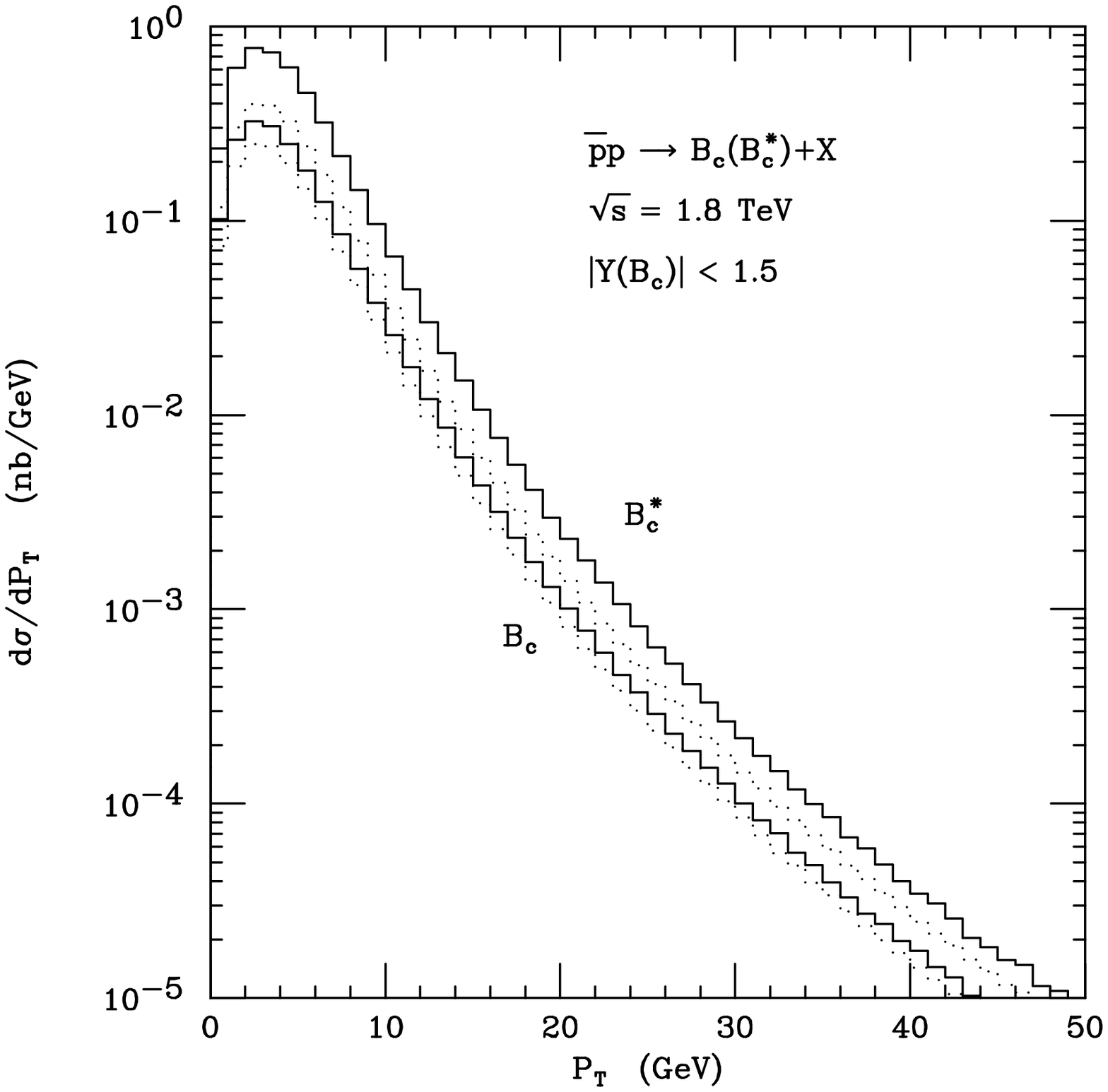}
\end{center}
\vspace*{13pt}
\fcaption{The $P_T$ distributions for $B_c$ and $B_c^*$ meson 
production at the Tevatron energy $\sqrt{{s}}=$ 1.8 TeV. The solid  
and the doted lines correspond to the full $\alpha_s^4$ calculation 
and the fragmentation approximation, respectively. }
\end{figure}
From  Fig. 2, we see that for the $B_c$ meson 
the $P_T$ distributions for $P_T>5$ GeV are very similar for the full 
$\alpha_s^4$ calculations and for the fragmentation approximation.
This critical value of $P_T$ is  much smaller than that found above in the 
study of the subprocess.  However, for the $B_c^*$ meson, the result 
predicted by the full $\alpha_s^4$ calculation in Fig. 2  is a factor 
1.5 to 2.0 greater than the fragmentation calculation over a much 
larger range of $P_T$, more consistent with what was found in the study
of the subprocess. Therefore, it is difficult to decide from only 
the $P_T$ distributions where the fragmentation approximation is reliable 
for the hadronic production of the $B_c$ and $B_c^*$ mesons. 
The use of $P_T$ distribution alone can be misleading.

To clarify this issue, we introduce the distribution
\begin{equation}
C(z) \displaystyle = {d \sigma \over d z } = \int dx_1 dx_2 g(x_1,\mu) 
g(x_2,\mu) {d\hat{\sigma}(\sqrt{\hat s},\mu) \over d z},
\end{equation}
where $\displaystyle z\equiv {2(k_1+k_2)\cdot p / {\hat{s}}}$
and $g(x_i,\mu)$ is the gluon distribution function.
The distribution $C(z)$ provides a sensitive means to investigate
the dynamics of the production process and the fragmentation approximation. 
Clearly, if the fragmentation approximation is valid, 
$\displaystyle {d\hat{\sigma}(\sqrt{\hat s},\mu ) / d z}$ 
can be factorized  as 
\begin{equation}
\displaystyle {d\hat{\sigma}(\sqrt{\hat s},\mu) \over d z} =
\sum_i \hat{\sigma}_{gg\to Q_i\bar{Q}_i}\otimes  
D_{Q_i\to B_c}(z,\mu).
\end{equation}
where $D_{Q_i\to B_c}(z,\mu)$ are the usual fragmentation functions
and $\hat{\sigma}_{gg\to Q_i\bar{Q}_i}$ is the gluon fusion subprocess
cross section for production of the heavy quark pair $Q_i\bar{Q}_i$.
In this approximation, the integrals over $x_1$ and $x_2$ can be performed, the 
fragmentation function can be factored out, and $C(z)$ 
is simply proportional to a sum of the usual fragmentation functions
which is insensitive to the parton distribution functions 
and to the kinematic cuts.
However, if the distribution $C(z)$ is quite different from the 
fragmentation functions, it is an 
indication that the fragmentation approximation is not valid. Therefore, 
comparing $C(z)$ calculated in the fragmentation approximation with 
the full order $\alpha_s^4$ calculation, provides a quantitative criterion 
to judge the validity of the fragmentation approximation. 

In Fig 3, the  distribution $C(z)$ calculated in the 
fragmentation approximation is compared with the full order $\alpha_s^4$
calculation for $B_c$ and $B_c^*$ meson production with a cut of $P_T>10$ GeV
(Fig. 3a) and also with cuts of $P_T>20$ GeV and $P_T>30$ GeV (Fig. 3b).
\begin{figure}[htbp]
\vspace*{13pt}
\begin{center}
\leavevmode
\epsfysize=2.3 truein
\epsfbox{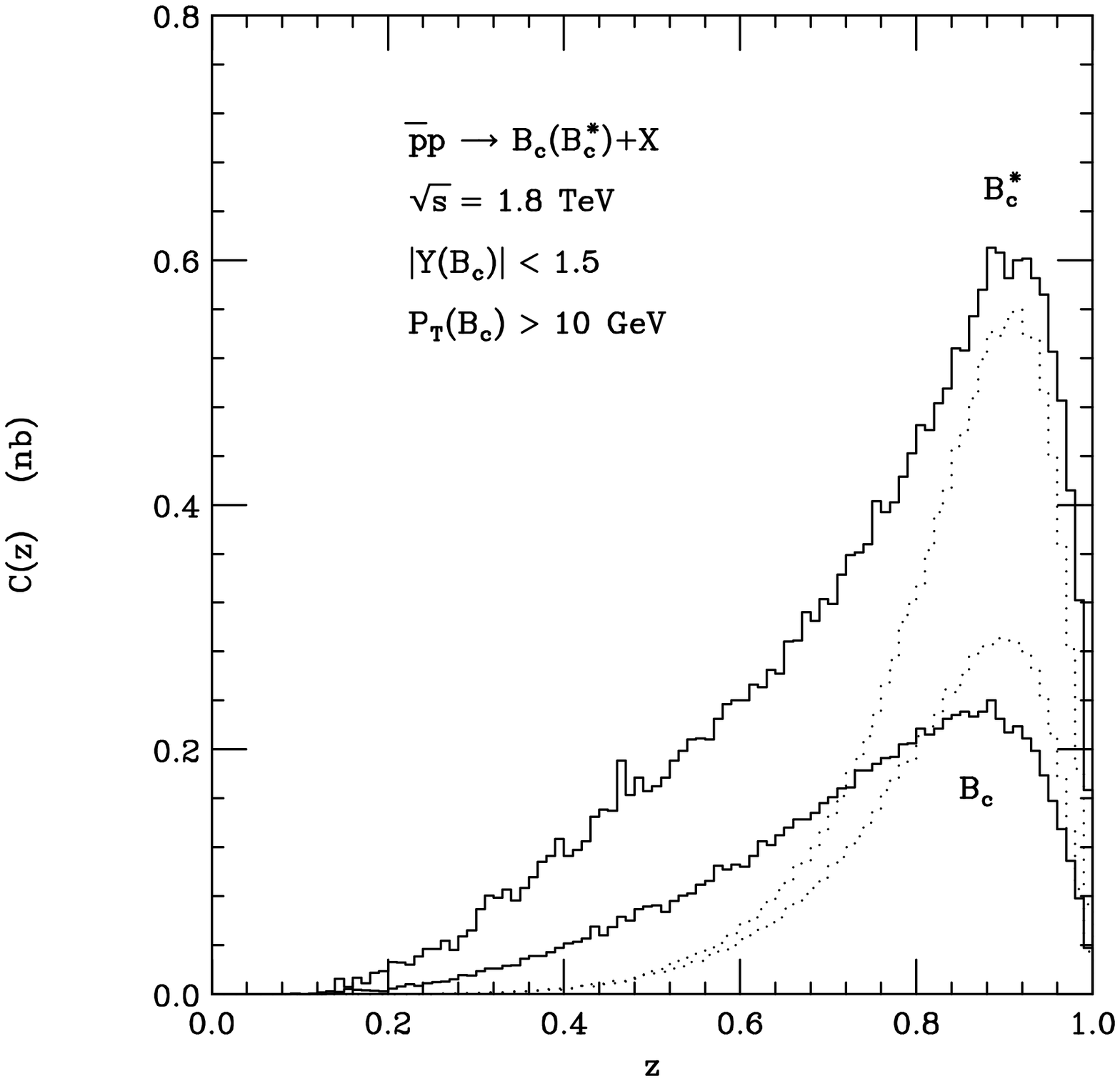} 
\hfill
\epsfysize=2.3 truein
\epsfbox{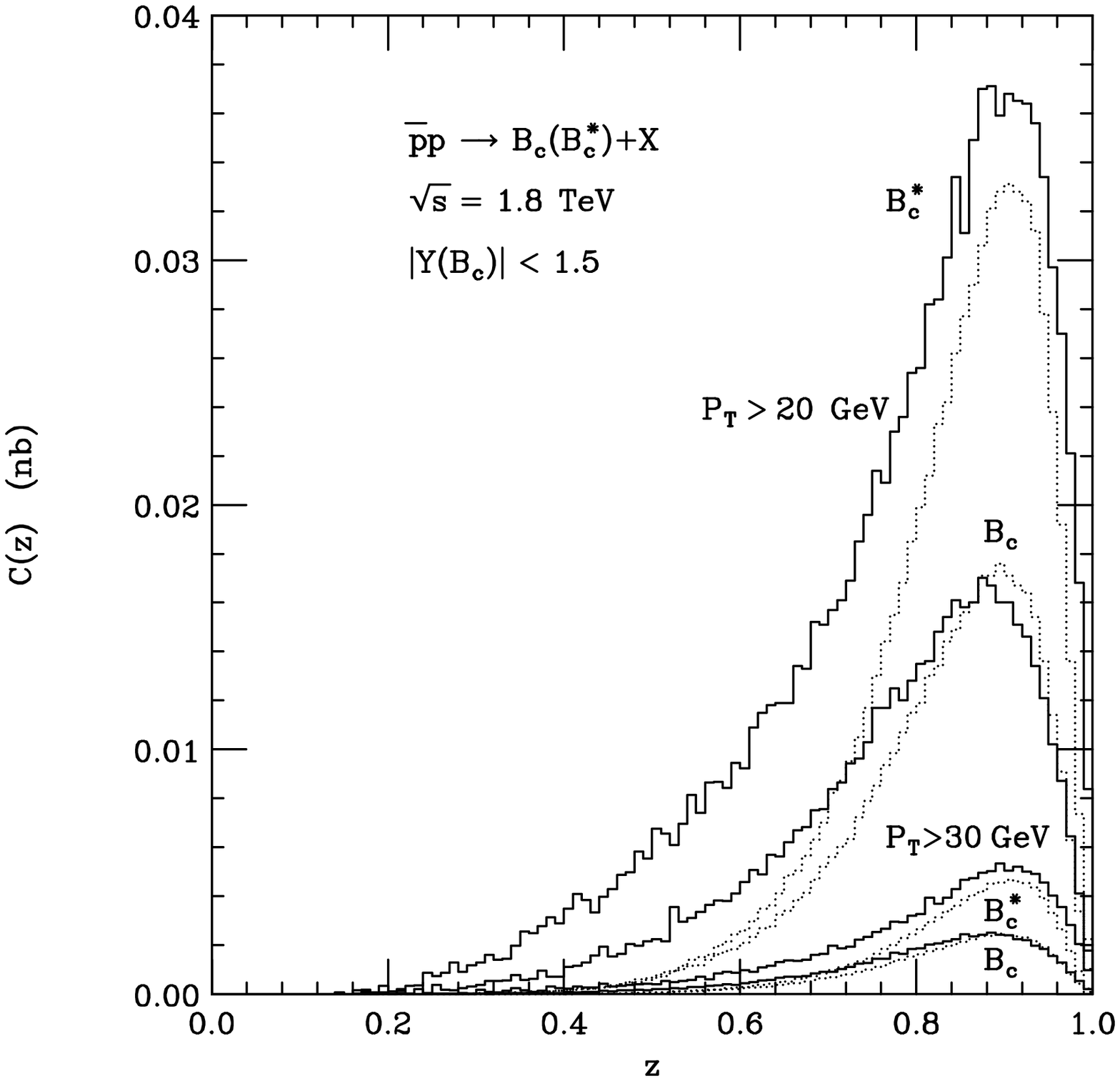}
\end{center}
\vspace*{13pt}
\fcaption{ The $z$ distributions $C(z)$ of the $B_c$ and $B_c^*$ 
at the Tevatron energy $\sqrt{s}=1.8$ TeV. The solid lines are the
full $\alpha_s^4$ calculation and the doted lines are the fragmentation
approximation (a) with the cut $P_T> 10$ GeV  and  (b) with the cuts 
$P_T> 20$ GeV and $P_T> 30$ GeV. }
\end{figure}
The $z$ distribution $C(z)$ is sensitive to 
the smallest $P_T$ region for a given 
$P_T$ cut because the $P_T$ distributions of the $B_c$ and $B_c^*$ mesons 
decrease very rapidly with increasing $P_T$.  From Fig. 3, some general 
features are evident. For the $B_c$ meson, 
with smaller $P_T$ cuts, the  distribution $C(z)$ is overestimated 
in the higher $z$ region  by the fragmentation approximation, 
while it is underestimated in the lower $z$ region, 
however, after integration over $z$, the result is similar to the full 
$\alpha_s^4$ calculation. For the $B_c^*$ meson, even for 
the largest $P_T$ cut, the distribution $C(z)$   is underestimated the 
fragmentation approximation at all values of $z$ and, 
after integration over $z$, the result is definitely smaller than the 
full $\alpha_s^4$ calculation. This feature explains why the $P_T$ 
distributions shown in Fig. 2 are similar for the $B_c$ meson even down 
to $P_T\sim M_{B_c} $ but are different for the $B_c^*$ meson. 
This also shows that it is simply fortuitous that the $P_T$ distribution 
of the $B_c$ calculated in the fragmentation approximation is similar to 
that from the full $\alpha_s^4$ calculation for $P_T$ below this value,
particularly down to $P_T\sim M_{B_c}$.  It is also clear that when 
$P_T$ is increased  the distributions become closer.
As shown in Fig. 3b, when $P_T$ is as large as 30 GeV the curves calculated
in the fragmentation approximation are quite close to the full $\alpha_s^4$
calculation. This indicates that the fragmentation 
approximation is valid in the large $P_T$ region, as expected. 
We emphasize here that the difference between the fragmentation 
approximation and full calculation is not universal, but is 
process--dependent. 

 Finally, we examine the ratio of cross sections for $B_c^*$ production
calculated in the fragmentation approximation with the results of the full 
$\alpha_s^4$ calculation. As discussed above, for $B_c^*$ meson production 
the fragmentation approximation always underestimates the full $\alpha_s^4$
result. 
The deviation from the full calculation for the $B_c^*$ meson can be used 
as a criterion to test the validity of the fragmentation approximation. The 
results for the total cross section $\sigma(P_T> P_{T \min})$ for various 
$P_T$ cuts  are listed in  Table I. 
Taking agreement within $30\%$ as the criterion for the validity of the 
fragmentation approximation we also see that $P_T$ should exceed about 
30 GeV, a value considerably larger than the heavy quark masses. 
We note that this conclusion is rather insensitive to the choice of
the energy scale $\mu$ and the parton distribution functions.

\begin{table}[htbp]
\tcaption{ Total cross sections 
$\sigma(P_{T\,B_c}> P_{T\,min})$  in $nb$ for hadronic 
production of the $B_c$ and the $B_c^*$ mesons predicted  by the 
$\alpha_s^4$ calculation and the fragmentation approximation assuming
various $P_T$ cuts and $|Y|<1.5$. 
The CTEQ3M parton distribution functions were used and the values
$f_{B_c}=480$ MeV, $m_c=1.5$ GeV, $m_b=4.9$
GeV, and $M_{B_c}=6.4$ GeV.}
\centerline{\footnotesize\smalllineskip 
\begin{tabular}{l|cccccc}
\multicolumn{7}{c}{}\\
\hline\hline 
 $P_{T\,min}$ (GeV) ~&  {0} & {5} 
 & {10}& {15}& {20}&  
    {30}  \\ \hline 
 $\sigma_{B_c}(\alpha_s^4)$ & 1.8 & 0.57 & 0.087 & 0.018 & $4.8\times10^{-3}$
          & $6.3\times10^{-4}$ \\
 $\sigma_{B_c}(\,frag.\,) $ & 1.4 & 0.47 & 0.071 & 0.014 & $4.0\times10^{-3}$
          & $5.3\times10^{-4}$  \\
 $\sigma_{B_c^*}(\alpha_s^4)$ & 4.4 & 1.4 & 0.22 & 0.041 & $1.1\times10^{-2}$
          & $1.3\times10^{-3}$ \\
 $\sigma_{B_c^*}(\,frag.\,) $ & 2.3 & 0.78 & 0.12 &0.025 & $6.8\times10^{-3}$
          & $9.2\times10^{-4}$  \\[2mm]
 $\displaystyle
  {\sigma_{B_c^*}(\,frag.\,) \over  \sigma_{B_c^*}(\alpha_s^4) }$ 
        & 0.52 & 0.55 & 0.56 & 0.61 & 0.63  & 0.70 
\\ [4mm]
\hline \hline \end{tabular} }
\end{table}

In summary, from the study of 
both the parton subprocess and the hadronic process, we can conclude that the 
fragmentation mechanism dominates when and only when $P_T\gg M_{B_c}$. 
This conclusion is independent of the choice of the energy scales and the 
parton distribution functions. It is only fortuitous that the $P_T$ 
distribution of the $B_c$ in the fragmentation approximation is similar to 
that of the full calculation for $P_T$ as low as $M_{B_c}$. 

\vspace*{1pt}\textlineskip	
\section{ Conclusion } 
\vspace*{-0.5pt}
\noindent
Both the spectroscopy of $\bar{b} c$ system and the decays of the 
$B_c$ mesons have been widely 
studied\cite{17,52,53,54}. The excited states below the threshold will 
decay to the ground state $B_c$ by emitting the photon(s) or $\pi$ mesons.  
The golden channel to detect the $B_c$ meson is  $B_c\longrightarrow J/\psi
+\pi (\rho)$. However, the branching  ratio is quite small, around 0.2. 
The exclusive semileptonic decay mode $B_c\longrightarrow J/\psi +l+\nu_l$  
has a relatively larger branch ratio, but there is ``missing energy''.
Comparative studies indicate that 
while fragmentation  approximation works very  well for the production
of the $B_c$ and $B_c^*$ in high energy 
$e^+e^-$ collisions, the condition for the applicability of the 
fragmentation approximation in hadronic collisions is that the transverse 
momentum $P_T$ of the  produced $B_c$ and $B_c^*$ be much larger than the 
mass of the $B_c$  meson; i.e., $P_T\gg M_{B_c}$. 
 Theoretical studies for the production of the $B_c$ and $B_c^*$ mesons 
indicate that an observable  number of $B_c$ and $B_c^*$ events can 
be produced at LEP I and at the Tevatron. 
 With the  discovery of the $B_c$  in the near
future, a new page in heavy flavor physics will be opened and 
 the study of the $B_c$ meson  will  no longer be academic.

\nonumsection{Acknowledgements}
\noindent
This research was done in collaboration with Chao-Hsi Chang and Robert J. Oakes.
I would like to thank Eric Braaten  for reading and improving 
this manuscript. 
We acknowledge the support of the U.S. Department of Energy, 
Division of High Energy
Physics, under Grant DE-FG02-91-ER40684 and the National Nature
Science Foundation of China.

\end{document}
\end